\documentstyle[preprint,prl,aps]{revtex}

\begin{document}

\preprint{DUKE-TH-96-133}

\draft

\title{Classical Lattice Gauge Fields with Hard Thermal Loops}

\author{C.~R. Hu\footnote{chaoran@phy.duke.edu}
        and B. M\"uller\footnote{muller@phy.duke.edu}}

\address{Department of Physics, Duke University, \\
Durham, North Carolina 27708--0305}

\date{\today}

\maketitle

\begin{abstract}

We propose a formulation of the long-distance dynamics of gauge 
theories at finite temperature on a lattice in Minkowski space, 
including the effects of hard thermal loops on the dynamics of the 
long wavelength modes.  Our approach is based on the dual classical 
limits of quantum fields as waves and particles in the infrared and 
ultraviolet limits, respectively. It exhibits manifest invariance 
under space-dependent lattice gauge transformations and conserves 
Gauss' law.

\end{abstract}


In the past few years great efforts have been made in perturbation
theory to calculate transport properties of thermal gauge fields from 
the low-energy effective action including the contribution of hard 
thermal loops \cite{BP90,TW90,Nair,BI,Thoma}.  Unfortunately, 
perturbative calculations fail in some important cases such as the 
damping of a traveling mode in the QCD plasma \cite{TG91}, color 
conductivity \cite{SG93}, and winding number diffusion \cite{OP93} 
due to the presence of singularities associated with the static 
magnetic gauge sector.  This difficulty has motivated numerical 
simulations of the dynamics of classical gauge fields in Minkowski 
space \cite{AAPS91,MT92,BGMT94,AK95}, which were based on the 
Hamiltonian
formulation of gauge theory on a spatial lattice \cite{KS75}.

These studies have been criticized \cite{BMS95,ASY96}, because they do 
not properly account for the influence of hard thermal loops which 
modifies
the dynamics of the long-distance modes of the gauge field.  For scalar 
field theories there exists a straightforward remedy \cite{BMS95,CM96}. 
One introduces a momentum cut-off $k_c$ and describes the influence of 
the high-momentum modes, which are essentially quantum mechanical, in 
perturbation theory. The dynamics of the long-distance modes then 
becomes 
dissipative and noisy \cite{CM96,BLL96}.  In the case of gauge theories, 
gauge invariance dictates the use of a lattice discretization  
eliminating modes with eigenvalues of the kinetic momentum operator 
$(-i\mbox{\boldmath $\nabla$}-g\mbox{\boldmath $A$})$ larger than $\pi/a$, 
where $a$ is the lattice spacing.  
However, the construction of the appropriate lattice action for the 
soft fields involving hard thermal loops encounters technical 
difficulties \cite{BMS95}.

Recently, Huet and Son \cite{HS96} have argued that it is possible to
construct an effective classical dynamics for quasi-static, long
wavelength modes of the gauge field containing dissipative and
stochastic terms.  The resulting equations are valid in the limit
$\omega \ll k\le g^2T$ where $\omega$ is the frequency and $k$ the
wave vector.  Even in this limit the noise term is strongly nonlocal
and no efficient numerical treatment was proposed by the authors.  In
the general case, for $\omega,k\ll T$, the hard thermal loop action
contains an infinite number of spatially and temporally nonlocal,
dissipative and stochastic vertices, which are difficult to treat
numerically.

We propose to circumvent these difficulties by representing the hard
thermal modes as classical colored particles propagating in the
background of the soft gauge fields.  Explicitly treating these modes
in terms of classical particles, rather than integrating them out,
leads to a set of {\it local} dynamical equations which can be
efficiently solved by numerical intergration after lattice
discretization.  Below we will construct a lattice version of these
equations that is gauge invariant and conserves Gauss' law.  Because
the dynamics of the hard modes is treated explicitly in this
formulation, it is not necessary to assume that they are thermally
populated.  One can, as well, consider dynamical situations far off
equilibrium, where the density matrix of hard modes of the gauge field
is characterized by some large scale.  Such conditions are, for example, 
of interest in the context of equilibration processes occurring in 
high energy nuclear collisions.

The representation of the high momentum components of the gauge field
in terms of classical particles propagating in classical gauge fields
can be justified by the eikonal limit of the Yang-Mills equations.
Heinz \cite{Heinz85} and Kelly et al. \cite{MIT94} have shown that a
thermal ensemble of particles obeying Wong's equations \cite{Wong70}
generates the correct hard thermal loop action \cite{BP90,TW90} for soft 
gauge fields.  The proof is based on linear response theory, combined 
with the explicit gauge covariance of the classical equations.  
In the following we briefly review the continuum formulation of the
classical transport theory \cite{Heinz90} and then show how the 
equations can be implemented on a spatial lattice.

The classical transport theory for nonabelian gauge fields starts from
the Boltzmann equation
\begin{equation}
p^{\mu} \left[ {\partial\over\partial x^{\mu}} - gQ^a F_{\mu\nu}^a
{\partial\over\partial p_{\nu}} - gf_{abc} A_{\mu}^bQ^c
{\partial\over\partial Q^a}\right] f(x,p,Q) = C[f] \label{e1}
\end{equation}
together with the Yang-Mills equations
\begin{equation}
D_{\mu}F^{\mu\nu} = g \int [dpdQ] p^{\nu}Q f(x,p,Q) \equiv
j^{\nu}(x)~, \label{fields}
\end{equation}
where $f(x,p,Q)$ denotes the one-particle phase space distribution of 
classical
particles, $Q$ is the classical nonabelian charge carried by the 
hard gluons, and $D_{\mu}$ is the gauge covariant
derivative.  Note that $p$ represents the kinetic (not the
canonical) momentum of the particles and therefore is gauge
invariant.  

For our purposes, it is sufficient to consider the Vlasov limit,
neglecting the collision term $C[f]$.  The transport equation
(\ref{e1}) is then solved by an ensemble of test particles
\begin{equation}
f(x,p,Q) = \frac{1}{N_0} \sum_i 
  \delta \left(x-\xi_i\right)
  \delta \left( p-p_i\right) 
  \delta \left( Q-Q_i\right)~,
\label{e3}
\end{equation}
where $N_0$ is the total number of particles. The space, momentum, and 
color coordinates of the particles obey Wong's equations \cite{Wong70}:
\begin{eqnarray}
 \dot{\mbox{\boldmath $\xi$}}_i &=& \mbox{\boldmath $v$}_i~,
         \label{position}\\
 \dot{\mbox{\boldmath $p$}}_i   &=& g Q^a_i 
     \left[\mbox{\boldmath $E$}^a(\xi_i)+
     \mbox{\boldmath $v$}_i\times\mbox{\boldmath $B$}^a(\xi_i)\right]~, 
         \label{momentum}\\
 {\dot{Q}}_i &=& -ig v^{\mu}_i \left[A_{\mu}(\xi_i),~Q_i\right]~.
         \label{charge}
\end{eqnarray}
The index $i$ enumerates the particles and $\xi^{\mu}_i=(t,
\mbox{\boldmath $\xi$}_i)$, $v^{\mu}=(1,\mbox{\boldmath $v$})$ with 
$\mbox{\boldmath $v$}_i=\mbox{\boldmath $p$}_i/|\mbox{\boldmath $p$}_i|$ 
being the velocity of the $i$th particle. Note that $v^{\mu}$, as 
defined here, is not a Lorentz four-vector. A dot denotes a time derivative.
$A_{\mu}$ denotes the vector potential, and $\mbox{\boldmath $E$}, 
$\mbox{\boldmath $B$} are the color electric and magnetic fields,  
respectively. The right-hand side of (\ref{momentum}) is a generalization 
of the electrodynamical Lorentz force.  Wong's equations can be written 
in manifestly covariant form, but here we have chosen a representation
that is convenient for the numerical implementation in the context of a 
Hamiltonian lattice gauge theory.

The Hamiltonian equations of motion for the gauge field and Gauss' law 
are obtained from (\ref{fields})  as the components $\nu=1,2,3$ and $\nu=0$, 
respectively. The charge current $j^{\mu}=(\rho,\mbox{\boldmath $j$})$ 
is defined as
\begin{eqnarray}
  j^{\mu}(x) &=& g\sum_i Q_i v^{\mu}_i
             \delta^3(\mbox{\boldmath $x$}-\mbox{\boldmath $\xi$}_i)~. 
      \label{current}
\end{eqnarray}
 
Since {\boldmath $E$}, {\boldmath $B$}, and $Q$ transform covariantly 
under gauge transformations, it is easy to see that (\ref{position}) 
and (\ref{momentum}) are gauge invariant.  The gauge covariance property 
of (\ref{charge}) is best demonstrated by first recognizing that 
(\ref{charge}) has the following formal solution (omitting the index $i$):
\begin{equation}
 Q(t)={\cal U}(t,0)Q(0){\cal U}^{\dagger}(t,0)~, \label{rotate}
\end{equation}
where ${\cal U}(t,0)$ is the parallel transport operator along the 
particle's world line:
\begin{equation}
 {\cal U}(t,0) = {\rm T} \exp\left[-ig\int_0^tdt^{\prime}
\frac{d\xi^{\mu}}{dt^{\prime}}A_{\mu}(\xi)\right]~.
   \label{transportor}
\end{equation}
${\cal U}$ obeys the equation
\begin{equation}
 \frac{d{\xi}^{\mu}}{dt}D_{\mu}(\xi){\cal U}(t,0)=0~,\quad {\cal U}(0,0)=1~.
   \label{transport}
\end{equation}
Under a gauge transformation $G$, ${\cal U}(t)$ 
transforms as ${\cal U}(t,0) \rightarrow G(x(t)){\cal U}(t,0)G^{\dagger}
(x(0))$. Hence (\ref{rotate}) is gauge covariant, and so is (\ref{charge}). 
It is worth noting that both (\ref{charge}) and (\ref{rotate}) conserve the
magnitude of the charges: 
\begin{equation}
 \frac{d}{dt}\sum_a Q^a Q^a=0~.
\end{equation}
As the charged particle travels in the background field, its charge 
rotates in color space with an angular velocity $\omega^a = 
g v^{\mu}A^a_{\mu}(\xi)$, consistent with the notion of the gauge field 
$A^{\mu}$ as the connection on a curved manifold on which
the charge of a moving particle undergoes parallel transport.

We now describe how the equations
(\ref{fields},\ref{position},\ref{momentum},\ref{charge}) can
be solved numerically. We discretize (\ref{fields}) on a lattice with 
lattice spacing $a$.  Then the soft 
gauge fields are restricted to modes with $k<k_{\rm c}=\pi/a$. To avoid 
double-counting of modes the hard particles will be restricted to kinetic 
momenta $p>k_c$. The precise connection between $k_c$ and $a$ can be
established by the requirement that the plasmon mass takes on its 
correct value found in the continuum theory: $\omega_p =
{\sqrt{N_c}gT}/3$. 
In the proposed simulation, we are mainly interested in modeling 
physics at the energy scale of $g^2 T$ and the dominant scattering process 
between hard particles happens at the scale of $g T$. In order to
get these right on the lattice, we must set up the simulation in such a
way that the following relation between the length scales is satisfied:
\begin{equation}
a < (g T)^{-1} < (g^2 T)^{-1} < N a ~,
\end{equation}
where $N a$ is the lattice size.

For discretization of the fields, we use the Kogut-Susskind (KS) model, 
which is summarized below. Improved lattice Hamiltonians are also possible 
\cite{GM96}.  In the KS scheme, one chooses the temporal gauge $A_0=0$ 
and expresses the gauge field in terms of variables $U_{x,i}$
associated with the link $(x,i)$ directed from a site $x$ to its nearest 
neighbor $x+i$:
\begin{equation}
  U_{x,i} = \exp \left[ -iga A_i^a (x) \,\tau^a/2 \right] = 
            U^{\dagger}_{x+i,-i}~.  \label{link}
\end{equation}
Under a gauge transformation $G$, the link variables transform as
\begin{equation}
 U_{x,i}\rightarrow G(x)U_{x,i}G^{\dagger}(x+i)~.
 \label{link_trans}
\end{equation}
A plaquette variable is defined as the product of four link variables 
associated with the sides of an elementary plaquette $(x,ij)$:
\begin{equation}
   U_{x,ij} = U_{x,i} \, U_{x+i,j} \, U_{x+i+j,-i} \, U_{x+j,-j}~.
  \label{plaquette}
\end{equation}
The links are directed and hence the plaquettes are oriented.
The electric and the magnetic fields are defined as
\begin{eqnarray}
 E_{x,i} &=& \frac{1}{iga} \dot{U}_{x,i}U^{\dagger}_{x,i}~, 
   \label{E_field} \\
 B_{x,k} &=& \frac{1}{4iga^2}\epsilon_{kij}(U^{\dagger}_{x,ij}-U_{x,ij})~,
   \label{B_field}
\end{eqnarray}
where $E_{x,i}$ is associated with the link $(x,i)$ and $B_{x,k}$ with the 
plaquette $(x,ij)$.  Both $E_{x,i}$ and $B_{x,k}$ transform covariantly
under a gauge transformation $G(x)$.

In the spirit of the Hamiltonian formalism, we choose $U_{x,i}$ and 
$E_{x,i}$ as the basic dynamic variables.  They obey 
the following equations of motion:
\begin{eqnarray}
 \dot{U}_{x,i} &=& igaE_{x,i}U_{x,i}~, 
  \label{U-dot} \\
 \dot{E}_{x,i} &=& 
       \frac{1}{2iga^3}\sum_j(U^{\dagger}_{x,ij}-U_{x,ij}) - j_{x,i}~,
  \label{E-dot}
\end{eqnarray}
and are subject to the constraint of Gauss' law:
\begin{equation}
 \frac{1}{a}
 \sum_i \left[E_{x,i}-U^{\dagger}_{x-i,i}E_{x-i,i}U_{x-i,i}\right]-\rho_x=0~.
  \label{gauss}
\end{equation} 

In order to define the charge current four-vector $j_x^{\mu}$ on the
 lattice, we take each site $x$ as the center of a cubic
cell $C_x$ of size $a^3$.  The color charge of every particle in $C_x$
will be counted as contribution to the charge density $\rho_x$.  Any
particles entering or leaving the box during a given time step $\Delta
t$ will contribute to the component of the color current normal to the
face of the cube that is penetrated \cite{thanks}:
\begin{equation}
\rho_x(t) = {g\over a^3} \sum_{k\in C_x} Q_k(t) \label{e21}~,
\end{equation}
\begin{equation}
j_{x,i}(t+{\Delta t}/2) = {g\over a^2\Delta t} \sum_{k\in C_x}^{(i)}
\left[Q_k(t)- Q_k(t+\Delta t)\right]~, \label{e22}
\end{equation}
where the notation in (\ref{e22}) indicates that only particles
entering or leaving along the link connecting $x$ and $x+i$ are counted.  

The variables to be advanced in time according to their equations of
motion are $U_{x,i}$, $E_{x,i}$, $\mbox{\boldmath $\xi$}_k$, 
$\mbox{\boldmath $p$}_k$, and $Q_k$. This can be achieved with a
modified leapfrog algorithm that conserves energy and satisfies Gauss' 
law.  It is most convenient to choose an update scheme in which the link 
variables $U_{x,i}$ are defined at half integer time steps while $E_{x,i}$, 
$\mbox{\boldmath $\xi$}_k$, $\mbox{\boldmath $p$}_k$, and $Q_k$ are defined
at integer time steps \cite{22}.  We only briefly discuss the momentum
update here.

The momentum gets contributions from both the electric and the magnetic 
fields. In the same spirit as our definition of the current $j_{x,i}$ 
in (\ref{e22}), only those particles which move from one cell to another 
during one time step obtain a momentum kick by the electric field:
\begin{equation}
p_{k,i}(t+\Delta t)= 
 p_{k,i}(t) + g Q^a_k(t)E^a_{x,i}(t)
\left[\frac{a}{v_{k,i}}
   \left( 1-\frac{1}{2a^2}
   \frac{{\rm tr}\left[Q_k(t)Q_k(t)\right]}
    {{\rm tr}\left[Q_k(t)E_{x,i}(t)\right]}
   \right)\right]~,
\label{e25}
\end{equation}
if $k\in C_x$ at time $t$ and $k\in C_{x+i}$ at time $t+\Delta t$. The
expression in the brackets can be regarded as the effective time needed
for a particle to transit from one cell to its nearest neighbor.
This choice balances the change in the energy of the particle and the
change in the energy of the lattice fields incurred by the transition
of that particle from one cell to another.  The influence of the magnetic 
field on the particle momenta is updated continuously during each time
step. Such a modified leapfrog algorithm is stable for 
${\Delta t}/a \stackrel{<}{\sim} 0.1$.

Finally, we point out that the definition (\ref{e22})
ensures the conservation of the color current during each time step:
\begin{equation}
{1\over\Delta t} \left( \rho_x (t+\Delta t) - \rho_x(t)\right) +
{1\over a} \sum_i \left( j_{x,i} - U_{x-i,i}^{\dagger} j_{x-i,i}
U_{x-i,i} \right) = 0~, \label{e23}
\end{equation}
where the quantities in the sum are defined at $t+{\Delta t}/2$.
This, together with the update scheme described above, automatically
ensures the exact conservation of the left-hand side of Gauss' law
(\ref{gauss}) from one time step to the next.
An important feature of our implementation is  
that all equations transform correctly under spatial gauge 
transformations.

The dynamical evolution of (soft) gauge field and (hard) particles
described by the equations of motion is essentially classical. 
However, we need to incorporate certain quantum features into 
the simulation, which are embodied in the initial conditions.  There
are two places where quantum physics enters.

First, while the energy and the momentum of a charged particle are
continuous classical variables, the nonabelian charge $Q^a$ contains 
a factor of $\hbar$ and has a fixed magnitude. In the classical limit
the color charge of a gauge boson rotates in color space like a 
three-vector with fixed length just as the spin of an electron rotates
in a magnetic field.  In the quantum case, the charges $Q^a$ are 
$q$-numbers and obey the SU(2) Lie algebra:
\begin{equation}
 [Q^a,~Q^b]=i\hbar \epsilon^{abc}Q^c~.
\end{equation}
In the semiclassical limit, we treat them as $c$-numbers but retain their
magnitude as proportional to the Casimir operator in the adjoint 
representation \cite{Heinz85}:
\begin{equation}
 \sum_a Q^aQ^a = 2\hbar^2~,
\end{equation}
which is conserved by (\ref{charge}). For particles in the fundamental
representation of SU(2), such as fermions, the right-hand side would
be replaced by $3\hbar^2/4$.

Second, while the dynamics is classical, we require that both the particles
representing hard thermal gluons and the fields obey Bose statistics. 
To illustrate this point, we consider the initialization of the particle 
ensemble at a certain temperature $T$. According to Bose statistics, the 
particles should be initialized with the distribution
\begin{equation}
 f(\mbox{\boldmath $x$, $p$}, Q)=
  \frac{\theta(p-\hbar k_{\rm c})}
  {\displaystyle{e^{\beta{\epsilon(p)}}-1}}
  \delta (Q^2-2\hbar^2)~,
\end{equation}
where $\epsilon(p)$ is the energy of a particle with momentum {\boldmath $p$}. 
The initial particle number density is then given by
\begin{equation}
 n(\mbox{\boldmath $x$})
           ={2N_c\over {h^3}}\int\frac{d^3\mbox{\boldmath $p$}}
            {\displaystyle{e^{\beta{\epsilon(p)}}-1}}
            \theta(p-\hbar k_{\rm c})~,
\end{equation}
where the factor $2N_c$ counts the spin and color degeneracies.
The linear response of the distribution $f(\mbox{\boldmath $x$, $p$}, Q)$ 
to the classical field then reproduces the HTL polarization function
\cite{Heinz85,MIT94}.

{\it Acknowledgements}: We thank U. Heinz for useful comments on the 
manuscript and G. Moore for valuable discussions and for advice on
formulating the lattice algorithm.  BM thanks C. Greiner, S. Leupold, 
and M. Thoma for discussions.  This work was supported in part by the 
U.S.~Department of Energy (Grant No. DE-FG02-96ER40945).



\begin{references}
\bibitem{22}
Details of this algorithm will be discussed in a forthcoming
manuscript where we will present first numerical results.
\end{references}
\end{document}